# METRICGAN-U: UNSUPERVISED SPEECH ENHANCEMENT/ DEREVERBERATION BASED ONLY ON NOISY/ REVERBERATED SPEECH

*Szu-Wei Fu[1*], Cheng Yu[1], Kuo-Hsuan Hung[1], Mirco Ravanelli[2], Yu Tsao[1]*

[1] Research Center for Information Technology Innovation, Academia Sinica, Taipei, Taiwan
[2] Mila-Quebec AI Institute, Montreal, Canada

## ABSTRACT

Most of the deep learning-based speech enhancement models are learned in a supervised manner, which implies that pairs of noisy and clean speech are required during training. Consequently, several noisy speeches recorded in daily life cannot be used to train the model. Although certain unsupervised learning frameworks have also been proposed to solve the "pair" constraint, they still require clean speech **or** noise for training. Therefore, in this paper, we propose MetricGAN-U, which stands for MetricGAN-unsupervised, to further release the constraint from conventional unsupervised learning. In MetricGAN-U, only noisy speech is required to train the model by optimizing non-intrusive speech quality metrics. The experimental results verified that MetricGAN-U outperforms baselines in both objective and subjective metrics.

## 1. INTRODUCTION

Recently, deep learning-based speech enhancement models have gained significant improvements compared to traditional methods [1-8]. However, the success is mainly based on a large amount of training data, which includes several different clean and noisy speech pairs. In general, the noisy speech is synthesized by adding clean speech with noise; hence, **both** clean speech and noise are required for model training. Because both are very difficult to obtain in daily life, they are usually recorded in an anechoic chamber. This implies that unlike noisy speech, a lot of time and effort is needed to collect them. Therefore, to solve this issue, in this study, we propose a speech enhancement/dereverberation framework whose training data are based only on noisy/reverberated speech.

In the field of speech enhancement, unsupervised training is usually defined as *pairs of noisy and clean speech signals that are **not** required during model training* [9]. This can be further divided into different levels of "unsupervised," according to what training data are actually needed, as follows:

1) *Level 1 of unsupervised learning (*clean speech **or** noise is needed*)*:

   Although noisy and clean speech pairs are not required, clean speech **or** noise may still be needed. Most unsupervised speech enhancement methods belong to this category. Bie *et al.* [9] used clean speech to first pre-train a variational auto-encoder and applied variational expectation-maximization to fine-tune the encoder part during inference. Another method to achieve non-parallel speech enhancement is through a cycle-consistent generative adversarial network (CycleGAN) [10, 11]. Through the framework of a GAN and cycle-consistent loss, only non-paired clean and noisy speech was used during training. Fujimura *et al.* [12] proposed noisy-target training (NyTT) by adding noise to noisy speech. The noise-added signal and original noisy speech are considered as the model input and target, respectively. In summary, although the above-mentioned methods do not require parallel data, clean speech or noise may still be needed.

2) *Level 2 of unsupervised learning (noisy speech is needed)*:

   In this study, we train a speech enhancement/ dereverberation model based on noisy/reverberated speech only. Because the proposed method follows the MetricGAN [13] framework, we call it MetricGAN-U, which is short for MetricGAN-unsupervised. The basic idea is to optimize a non-intrusive speech quality metric so that no clean speech is required during training. We selected DNSMOS [14] and speech-to-reverberation modulation energy ratio (SRMR) [15] as the quality metrics for speech enhancement and dereverberation, respectively. To foster reproducibility, MetricGAN-U is available within the SpeechBrain[1] [16].

3) *Level 3 of unsupervised learning (no training data is needed)*:

   At this level of unsupervised learning, no training data is required (because it does not even learn anything!). Most of these methods are based on traditional signal processing methods, such as the MMSE [17] and Wiener filter [18].

   **Note that the extra effort needed to go from level 3 to level 2 is not as large as level 2 to level 1, because noisy speech is very easy to obtain in our daily life.**

---

*Currently affiliated with Microsoft.

[1] Code is available at https://speechbrain.github.io/

## 2. INTRODUCTION TO METRICGAN

The training framework of MetricGAN [13] is very similar to a conventional generative adversarial network (GAN) [19], except that the goal of the discriminator is to mimic the behavior of the target evaluation function (e.g., perceptual evaluation of speech quality (PESQ) function [20]). The discriminator (surrogate of the metric of interest) is learned from raw metric scores by treating the target evaluation function as a black box. Hence, the surrogate can be used as a loss function for the generator (speech enhancement model) to optimize the target metric. We recently proposed an improved version of MetricGAN called MetricGAN+ [21], which includes some more advanced training techniques. We briefly introduce the training algorithm here.

Let $Q'(I)$ be a function that represents the target evaluation metric normalized between 0 and 1, where $I$ denotes the input of the metric. For example, for intrusive metrics such as PESQ, $I$ denotes a pair of enhanced speech, $G(x)$ (or noisy speech, $x$) that we want to evaluate, and its corresponding clean speech, $y$. To ensure that the discriminator network ($D$) behaves similar to $Q'$, the objective function of $D$ is

$$L_{D(\text{MetricGAN+})} = \mathbb{E}_{x,y}[(D(y,y) - Q'(y,y))^2 + (D(G(x),y) - Q'(G(x),y))^2 + (D(x,y) - Q'(x,y))^2] \quad (1)$$

The three terms were used to minimize the difference between $D(.)$ and $Q'(.)$ for clean, enhanced, and noisy speech, respectively. Note that, $Q'(y,y) = 1$, $0 \leq Q'(G(x),y) \leq 1$ and $0 \leq Q'(x,y) \leq 1$.

The training of the generator network ($G$) can completely rely on the adversarial loss

$$L_{G(\text{MetricGAN+})} = \mathbb{E}_x[(D(G(x),y) - s)^2] \quad (2)$$

where $s$ denotes the desired assigned score. For example, to generate clean speech, we can simply assign 1 to s.

## 3. UNSUPERVISED SPEECH ENHANCEMENT /DEREVERBERATION USING METRICGAN FRAMEWORK

### 3.1. MetricGAN-U

To achieve level 2 of unsupervised speech enhancement/dereverberation, the clean speech, $y$, used in (1) and (2) have to be removed. Therefore, the training of the discriminator network is modified as:

$$L_{D(\text{MetricGAN-U})} = \mathbb{E}_x[(D(G(x)) - Q'(G(x)))^2 + (D(x) - Q'(x))^2] \quad (3)$$

From (3), it is obvious that the target evaluation metric $Q'(.)$ must be nonintrusive (no clean reference is needed). Certain well-known non-intrusive speech quality metrics include ITU-T P.563 [22] (designed for 3.1 kHz narrow-band telephony applications), SRMR [15], and DNSMOS [14]. In this study, we apply DNSMOS as $Q'(.)$ for speech enhancement and SRMR for speech dereverberation. The overall training flow for MetricGAN-U is shown in Fig. 1.

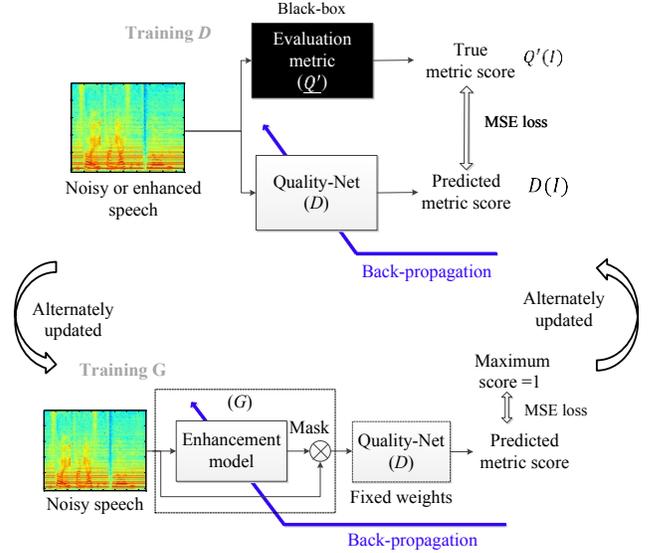

**Fig. 1.** Training flow of MetricGAN-U.

### 3.2. Introduction to DNSMOS and SRMR

DNSMOS is a neural network based quality estimation metric that can be used to evaluate different deep noise suppression (DNS) methods based on mean opinion score (MOS) estimates [14]. It is trained using ground truth human ratings obtained using ITU-T P.808 [23]. Although theoretically we can just directly concatenate the DNSMOS model after a speech enhancement model and use backpropagation to increase its score, the DNSMOS model is not publicly released, and hence, we can only obtain the scores through an API provided by the authors. Therefore, this also falls within the application scope of MetricGAN as a black-box metric optimization.

In contrast to DNSMOS, SRMR is a handcrafted metric design by speech experts. The basic idea of SRMR is based on the observation that reverberation generates high-frequency modulation energy. Hence, its definition is simply the ratio between the low-frequency modulation energy (speech component) and high-frequency one (reverberation component). For more details, please refer to [15].

## 4. EXPERIMENTS

### 4.1. Model Structure

The generator used in this experiment is a BLSTM [24] with two bidirectional LSTM layers, with 200 neurons each. The LSTM is followed by two fully connected layers, each with 300 LeakyReLU nodes and 257 sigmoid nodes for mask estimation, respectively. The discriminator herein is a CNN with four two-dimensional (2-D) convolutional layers with 15 filters and a kernel size of (5, 5). To handle the variable length input, a 2-D global average pooling layer was added such that the features could be fixed at 15 dimensions. Three fully connected layers were subsequently added, each with 50 and 10 LeakyReLU neurons, and 1 linear node.

**Table 1**. Comparison of MetricGAN-U with other methods for speech enhancement on the VoiceBank-DEMAND test set.

| | Unsupervised? | Need clean (c), noise (n), or noisy (N) for training? | PESQ | CSIG | CBAK | COVL | **DNSMOS** |
|---|---|---|---|---|---|---|---|
| Noisy | - | - | 1.97 | 3.35 | 2.44 | 2.63 | 3.048 |
| SEGAN [25] | ✗ | N(n+c) & c | 2.16 | 3.48 | 2.94 | 2.80 | - |
| BLSTM (MSE) | ✗ | N(n+c) & c | 2.71 | 3.94 | 3.28 | 3.32 | 3.367 |
| MMSE [17] | ✓ (Level 3) | No | 2.19 | 3.16 | 2.55 | 2.61 | 2.978 |
| Wiener [18] | ✓ (Level 3) | No | 2.22 | 3.23 | 2.68 | 2.67 | 3.050 |
| NyTT [12] | ✓ (Level 1) | N & n | 2.30 | 3.19 | **3.01** | 2.72 | - |
| **MetricGAN-U (half)** | ✓ (Level 2) | N | **2.45** | **3.47** | 2.63 | **2.91** | 2.891 |
| **MetricGAN-U (full)** | ✓ (Level 2) | N | 2.13 | 3.22 | 2.42 | 2.63 | **3.151** |

### 4.1. Speech enhancement
*4.1.1. Dataset*
To compare the proposed MetricGAN-U with other existing methods, we used the publicly available VoiceBank-DEMAND dataset [26]. The original training sets (11572 utterances) consisted of 28 speakers with four signal-to-noise ratios (SNRs) (15, 10, 5, and 0 dB). We randomly selected two speakers (p226 and p287) from this set to form a validation set (770 utterances). The test sets (824 utterances) consisted of two speakers with four SNRs (17.5, 12.5, 7.5, and 2.5 dB). Details of the data can be found in the original paper. We mainly evaluated the performance with the PESQ and DNSMOS scores. Although the other three metrics, CSIG, CBAK, and COVL, predict the MOS of the signal distortion, background noise interference, and overall speech quality, respectively, they are all based on the PESQ [18] and may not be highly correlated to MOS [27] (e.g., CBAK is very sensitive to the loudness of speech).

*4.1.2. Experimental results*
Table 1 shows the results of the proposed MetricGAN-U with other supervised and unsupervised baselines. Note that, although the criterion to distinguish between supervised and unsupervised training here is based on whether clean and noisy training pairs are needed, the degree of unsupervised training can be further investigated. The "most" unsupervised methods are MMSE [17] and Wiener filter [18] where no training data is required. MetricGAN-U only requires noisy speech for training, and it can be real recordings (does not have to be synthetic). Finally, although noisy-target training (NyTT) [12] also does not require clean speech, noise is required to generate more noisy signals as model input (the original noisy speech is used as the target).

The difference between MetricGAN-U (half) and MetricGAN-U (full) was the number of training epochs. The former is based on the early stopping [28] strategy with the criterion that the average PESQ score on the validation set reaches a maximum, while the latter is trained with full 600 epochs. Fig. 2 shows the learning process of the DNSMOS scores on the validation set.

In Table 1, although the Wiener filter can improve the PESQ score, its DNSMOS score is almost the same as that of noisy speech. Similarly, our MetricGAN-U (half) can obtain the highest PESQ score among unsupervised methods, but its DNSMOS score is the lowest (it seems that the correlation between the two metrics does not always remain positive). On the other hand, MetricGAN-U (full) seems to achieve a good trade-off between all metrics. In summary, it can increase 0.16 and 0.103 for PESQ and DNSMOS, respectively (Note that, considering the DNSMOS score for the clean test set is only 3.567, an improvement of 0.103 is significant).

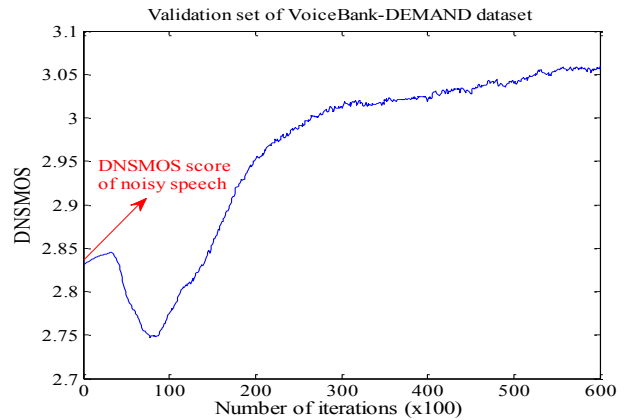

**Fig. 2.** Learning curve of MetricGAN-U on the validation set of VoiceBank-DEMAND dataset.

To evaluate the perceptual quality of the enhanced speech, we conducted subjective listening tests to compare the proposed MetricGAN-U (full) with Wiener filter and noisy speech (because it has been shown DNSMOS have higher correlation with MOS than PESQ [14], we selected MetricGAN-U (full) instead of MetricGAN-U (half)). 40 samples were randomly selected from the test set; therefore, there were a total of 40 × 3 (enhancement methods) = 120 utterances that each listener had to take. For each signal, the listener rated the overall quality based on signal distortion and noise intrusiveness, using a scale from 1 to 5. (e.g., 5 represents excellent quality, and 1 represents very poor quality). 15 listeners participated in the study. In Table 2, it can be observed that MetricGAN-U (full) is preferred over both the noisy and Wiener baseline (with p-value <0.001 and <0.05, respectively). An example of a spectrogram presentation is shown in Fig. 3. It can be found that MetricGAN-U (full) can remove more noise compared with the Wiener filter enhanced speech.

**Table 2**. Subjective evaluation results for speech enhancement.

|     | Noisy | Wiener | MetricGAN-U (full) |
|-----|-------|--------|--------------------|
| MOS | 3.083 | 3.387  | **3.556**          |

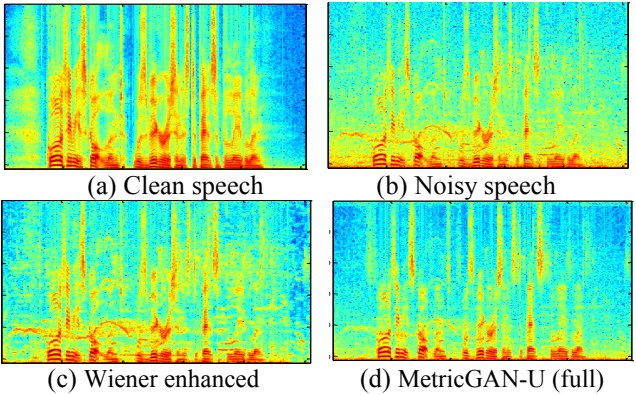

(a) Clean speech  (b) Noisy speech
(c) Wiener enhanced  (d) MetricGAN-U (full)

**Fig. 3**. Spectrograms of an example (2.5 dB) in the VoiceBank-DEMAND test set.

### 4.2. Speech dereverberation
*4.2.1. Dataset*
Because there is no publicly available free dataset for speech dereverberation, we prepared a dataset based on the clean speech from VoiceBank-DEMAND [26] (discard the noisy speech) and convolved them with the room impulse response (RIR) from OpenSLR[2]. We used the list[3] provided in the downloaded files to select 325 real RIR data for our dataset generation. Among them, 315 RIR were used to generate the training set, and another 10 RIR were randomly chosen for the test set. We applied the ***AddReverb*** function provided in the SpeechBrain [16] toolkit to generate the reverberated speech. The ***rir_scale_factor*** is randomly selected from {1.2, 1.1, 1.0, 0.9, 0.8} for the training set and {1.15, 1.05, 0.95, 0.85, 0.75} for the testing set. If $0 < $ ***rir_scale_factor*** $ < 1$, the impulse response is compressed (less reverb), whereas if ***rir_scale_factor*** $ > 1$, it is dilated (more reverb). Each clean utterance is convolved with only one RIR and ***rir_scale_factor***. As in the case of VoiceBank-DEMAND, two speakers (p226 and p287) were randomly chosen from the training set to form a validation set (770 utterances). The prepared dataset is called VoiceBank-SLR and can be publicly available[4] for a fair comparison.

*4.2.2. Experimental results*
In the experiment for speech dereverberation, MetricGAN-U was applied to maximize the SRMR score. Preliminary experiments show that the SRMR score can be easily increased to a very high level (i.e., larger than 10) by increasing the low-frequency modulation energy (speech component) and decreasing the high-frequency component (reverberation component). Because we only want to remove the reverberation component and keep the speech component unchanged, a self-reconstruction constraint term is added to the loss function of the generator:

$$L_{G(\text{MetricGAN-U(SRMR)})} = \mathbb{E}_x\left[\left(D(G(x)) - s\right)^2\right] + \alpha |G(x) - x|_2^2 \quad (4)$$

where $\alpha$ is set to 0.6, decided on the performance of the validation set. Table 3 shows that MetricGAN-U outperforms the well-known weighted prediction error (WPE) [29].

**Table 3**. Comparison of MetricGAN-U with other methods for speech dereverberation on the VoiceBank-SLR test set.

|              | Unsupervised? | PESQ | **SRMR** |
|--------------|---------------|------|----------|
| Reverb       | -             | 1.98 | 6.039    |
| BLSTM (MSE)  | ✗             | 2.35 | 8.039    |
| WPE [29]     | ✓             | 2.01 | 6.259    |
| **MetricGAN-U** | ✓          | **2.07** | **8.265** |

**Table 4**. Subjective evaluation results for speech dereverberation.

|     | Reverb | WPE   | MetricGAN-U |
|-----|--------|-------|-------------|
| MOS | 2.920  | 2.957 | **3.140**   |

We also conducted subjective listening tests on the dereverberation task to compare MetricGAN-U with WPE and reverb speech. 20 samples were randomly selected from the test set; therefore, there were a total of 20 × 3 (dereverb methods) = 60 utterances that each participant had to listen. For each signal, the listener rated the overall quality of dereverberation using a scale from 1 to 5. 15 listeners participated in the study. In Table 4, it can be observed that MetricGAN-U is preferred over both the reverb speech and WPE (with p-value <0.001 and <0.01, respectively).

### 5. DISCUSSION
Although it seems that there is a gap between the performance of supervised methods and the proposed unsupervised method, the difference can be easily reduced by increasing the amount of training data, which is very easy to obtain for level 2 of unsupervised learning. In addition, our method can also be used to boost the performance of supervised models by 1) applying the weights of the unsupervised model as the weight initialization for the supervised model or 2) as in [30], train the generator to maximize the discriminator score using real noisy data together with the synthetic one.

### 6. CONCLUSION
Compared to conventional unsupervised training, this paper proposed MetricGAN-U, which only requires noisy speech as training data. Because noisy speech is much easier to obtain than clean speech and noise, this significantly reduces the effort required for training data collection. By applying the proposed framework, a large amount of real noisy data can also be utilized during model training.

---
[2] http://www.openslr.org/resources/28/rirs_noises.zip
[3] In the path: RIRS_NOISES/real_rirs_isotropic_noises / rir_list
[4] https://bio-asplab.citi.sinica.edu.tw/Opensource.html#VB-SLR